\documentclass[
 twocolumn,
 nofootinbib,
 amsmath,
 amssymb,
 aps,
]{revtex4}

\usepackage{graphicx} 
\usepackage{dcolumn} 
\usepackage{bm} 

\usepackage{hyperref} 

\usepackage{xcolor}
\allowdisplaybreaks

\usepackage{mathtools}
\DeclareMathOperator*{\argmin}{arg\,min}

\usepackage{amsthm}
\theoremstyle{plain}
\newtheorem{theorem}{Theorem}

\theoremstyle{definition}

\theoremstyle{remark}

\usepackage{coffeestains}

\usepackage{xr}
\begin{document}

\preprint{APS/123-QED}

\title{Quantum natural gradient without monotonicity}

\author{Toi Sasaki}

\affiliation{
	Graduate School of Information Science and Technology,
	Hokkaido University, Sapporo, Hokkaido, 060-0814, Japan
}

\author{Hideyuki Miyahara}

\email{miyahara@ist.hokudai.ac.jp, hmiyahara512@gmail.com}

\thanks{corresponding author.}

\affiliation{
	Graduate School of Information Science and Technology,
	Hokkaido University, Sapporo, Hokkaido, 060-0814, Japan
}




%
%

\date{\today}

\begin{abstract}
	Natural gradient (NG) is an information-geometric optimization method that plays a crucial role, especially in the estimation of parameters for machine learning models like neural networks.
	To apply NG to quantum systems, the quantum natural gradient (QNG) was introduced and utilized for noisy intermediate-scale devices.
	Additionally, a mathematically equivalent approach to QNG, known as the stochastic reconfiguration method, has been implemented to enhance the performance of quantum Monte Carlo methods.
	It is worth noting that these methods are based on the symmetric logarithmic derivative (SLD) metric, which is one of the monotone metrics.
	So far, monotonicity has been believed to be a guiding principle to construct a geometry in physics.
	In this paper, we propose generalized QNG by removing the condition of monotonicity.
	Initially, we demonstrate that monotonicity is a crucial condition for conventional QNG to be optimal.
	Subsequently, we provide analytical and numerical evidence showing that non-monotone QNG outperforms conventional QNG based on the SLD metric in terms of convergence speed.
\end{abstract}


\maketitle



\section{Introduction}

Continuous optimization is prevalent in scientific and industrial problem-solving, with a plethora of methods proposed~\cite{Nocedal_001}.
Natural gradient (NG) stands out as a novel continuous optimization method rooted in information geometry (IG)~\cite{Amari_002, Amari_004, Amari_005}.
The field of machine learning has gained significant attention recently, with neural networks (NNs) holding paramount importance~\cite{Bishop_001, Murphy_002, Murphy_003}.
NG is frequently employed in the optimization of NNs, demonstrating numerical effectiveness~\cite{Martens_001}.

In the realm of physics, the era of quantum computing has dawned, prompting investigations into the hidden potential of noisy intermediate-scale quantum (NISQ) devices~\cite{Preskill_001}.
For NISQ devices, the parameter estimation of quantum circuits is imperative~\cite{Mitarai_001, Schuld_001, Miyahara_004}.
In addition to quantum circuit learning, estimating parameters of quantum states is essential in various contexts, including the variational Monte Carlo method~\cite{Scherer_001}, and tensor networks~\cite{Orus_001}.
Quantum natural gradient (QNG), introduced as the quantum counterpart of NG, has recently garnered attention for its promising performance in applications~\cite{Stokes_001, Koczor_001}.
A mathematically equivalent method to QNG, known in condensed matter physics, is the stochastic reconfiguration (SR) algorithm~\cite{Sorella_001, Sorella_002, Mazzola_001, Park_001, Xie_001}.
Furthermore, quantum metrics find applications in classical machine learning contexts~\cite{Lopatnikova_001, Miyahara_001, Miyahara_002, Miyahara_003}.
It is noteworthy that both conventional QNG and the SR method are grounded in the symmetric logarithmic derivative (SLD) metric, which is one of the monotone quantum Fisher metrics and a fundamental concept in quantum information geometry (QIG).

In this paper, we present a generalization of QNG based on QIG.
Monotonicity implies that information is not gained via completely positive and trace-preserving (CPTP) maps; so it has been believed to be a guiding principle for constructing a geometry in physics.
However, from the viewpoint of optimization, it has not been clear whether it is necessary.
Initially, we establish that monotonicity is a prerequisite for the SLD metric to optimize the speed of QNG.
Subsequently, we theoretically demonstrate that QNG utilizing non-monotone quantum Fisher metrics exhibits faster performance than its SLD-based counterpart.
Furthermore, in practical numerical simulations for both classical and quantum cases, the diagonal approximation of the Fisher metric is often employed to mitigate the computational costs associated with inverting the Fisher metric.
We show that even under this approximation, the aforementioned findings remain valid.
Lastly, we provide numerical simulations to illustrate that non-monotone quantum Fisher metrics outperform the SLD metric in terms of speed.

\section{Natural gradient} 

In the derivation of NG, the Fisher metric and Kullback-Leibler (KL) divergence play an essential role.
Then, we first explain the Fisher metric.
Assume that we have $p_\theta (\cdot)$, which is a probability distribution parameterized by $\theta \in \mathbb{R}^n$.
We define tangent vectors as $X \coloneqq \sum_{i=1}^n X^i \partial_i$, where $\partial_i \coloneqq \frac{\partial}{\partial \theta^i}$ for $i = 1, 2, \dots, n$.
Then, the Fisher metric between $X$ and $Y$ is defined as
\begin{align}
	g_{p_\theta (\cdot)} (X, Y) & \coloneqq \sum_{x=1}^N X_{p_\theta (\cdot)}^\mathrm{m} (x) Y_{p_\theta (\cdot)}^\mathrm{e} (x). \label{main_eq_def_Fisher_metric_001_001}
\end{align}
where the e- and the m-representations of $X$ at $p_\theta (\cdot)$ are given, respectively, by
\begin{subequations} \label{main_eq_classical_X_em_001_001}
	\begin{align}
		X_{p_\theta (\cdot)}^\mathrm{m} (\cdot) & \coloneqq X p_\theta (\cdot),                                           \\
		X_{p_\theta (\cdot)}^\mathrm{e} (\cdot) & \coloneqq X \ln p_\theta (\cdot). \label{main_eq_classical_X_e_001_001}
	\end{align}
\end{subequations}

Next, we turn our attention to classical rescaled R\'enyi divergence~\cite{Erven_001}.
When we explain NG only, KL divergence is sufficient; but, the purpose of this paper is to discuss QNG.
Then, we explain classical rescaled R\'enyi divergence, which is a generalized version of KL divergence.
Classical rescaled R\'enyi divergence is given by
\begin{align}
	D_\alpha (p_{\bar{\theta}} (\cdot) \| p_\theta (\cdot)) & \coloneqq \frac{1}{\alpha (\alpha - 1)} \ln \sum_{x=1}^N p_{\bar{\theta}}^\alpha (x) p_\theta^{1 - \alpha} (x). \label{main_eq_def_classical_rescaled_Renyi_divergence_001_001}
\end{align}
Note that ``rescaled" means the additional factor $1 / \alpha$ in Eq.~\eqref{main_eq_def_classical_rescaled_Renyi_divergence_001_001}.
Furthermore the rescaled R\'enyi divergence, Eq.~\eqref{main_eq_def_classical_rescaled_Renyi_divergence_001_001}, in the limit $\alpha \to 1$ corresponds to the KL divergence~\cite{Erven_001}:
\begin{align}
	\lim_{\alpha \to 1} D_\alpha (p_{\bar{\theta}} (\cdot) \| p_\theta (\cdot)) & = D_\mathrm{KL} (p_{\bar{\theta}} (\cdot) \| p_\theta (\cdot)).
\end{align}
Notably, Eqs.~\eqref{main_eq_def_Fisher_metric_001_001} and \eqref{main_eq_def_classical_rescaled_Renyi_divergence_001_001} satisfy the following relationship:
\begin{align}
	g_{p_\theta (\cdot)} (X, Y) & = \lim_{\bar{\theta} \to \theta} \bar{X} \bar{Y} D_\alpha (p_{\bar{\theta}} (\cdot) \| p_\theta (\cdot)), \label{main_eq_classical_Fisher_metric_001_001}
\end{align}
where $\bar{X}, \bar{Y}$ are the tangent vectors acting on $p_{\bar{\theta}} (\cdot)$.
The key observation is that the right-hand side of Eq.~\eqref{main_eq_classical_Fisher_metric_001_001} do not depend on $\alpha$.
As we will see later, the quantum counterpart of Eq.~\eqref{main_eq_def_classical_rescaled_Renyi_divergence_001_001} yields different metric contrarily to the classical limit, and this fact will play an important role in QNG.

We here review NG, which is one of the optimization methods for continuous variables motivated by IG~\cite{Amari_005}.
Let us consider the minimization problem of $L (\theta)$.
Then, the dynamics of NG is given by
\begin{subequations} \label{main_optimization_problem_NG_001_001}
	\begin{align}
		\theta_{\tau+1}          & = \theta_\tau + \Delta \theta (\epsilon),                                                                                                                                                           \\
		\Delta \theta (\epsilon) & = \argmin_{\Delta \theta: D_\alpha (p_{\theta_\tau + \Delta \theta} (\cdot) \| p_{\theta_\tau} (\cdot)) \le \epsilon} L (\theta_\tau + \Delta \theta), \label{main_optimization_problem_NG_001_012}
	\end{align}
\end{subequations}
where $\epsilon$ is a positive number.
The key point of Eq.~\eqref{main_optimization_problem_NG_001_001} is the constraint in Eq.~\eqref{main_optimization_problem_NG_001_012} since this term does not exist in conventional optimization methods such as the steepest decent method and Newton's method.
Note that in this paper, we focus on Eq.~\eqref{main_eq_def_classical_rescaled_Renyi_divergence_001_001}, but we do not need to limit ourselves to Eq.~\eqref{main_eq_def_classical_rescaled_Renyi_divergence_001_001}.
First, we transform Eq.~\eqref{main_optimization_problem_NG_001_012} by using the lowest-order approximation:
\begin{align}
	\Delta \theta (\epsilon) & \approx \argmin_{\Delta \theta: \frac{1}{2} \Delta \theta^\intercal G_\alpha (\theta_\tau) \Delta \theta \le \epsilon} \nabla_\theta L (\theta_\tau)^\intercal \Delta \theta, \label{main_optimization_problem_NG_002_001}
\end{align}
where $\nabla_\theta L (\theta) \coloneqq [\partial_1 L (\theta), \partial_2 L (\theta), \dots, \partial_n L (\theta)]^\intercal$, $\nabla_\theta L (\theta_\tau) \coloneqq \nabla_\theta L (\theta) |_{\theta = \theta_\tau}$, and $G_\alpha (\theta_\tau)$ is the matrix representation of the Fisher metric $g_{p_{\theta_\tau} (\cdot)} (\partial_i, \partial_j)$ induced by $D_\alpha (p_{\theta_\tau + \Delta \theta} (\cdot) \| p_{\theta_\tau} (\cdot))$.
To solve Eq.~\eqref{main_optimization_problem_NG_002_001}, we use the method of Lagrange's multiplier~\cite{Mizohata_001}.
Then the solution of Eq.~\eqref{main_optimization_problem_NG_002_001} is rewritten as
\begin{align}
	\Delta \theta (\epsilon) & = - \sqrt{\frac{2 \epsilon}{\nabla_\theta L (\theta_\tau)^\intercal G_\alpha^{-1} (\theta_\tau) \nabla_\theta L (\theta_\tau)}} G_\alpha^{-1} (\theta_\tau) \nabla_\theta L (\theta_\tau). \label{main_eq_update_theta_001_001}
\end{align}
Note that we have added a negative sign to decrease the value of $L (\theta)$ by this update.
In a practical usage of NG, we may use the following update rule instead of Eq.~\eqref{main_eq_update_theta_001_001}:
\begin{align}
	\Delta \theta (\epsilon) & = - \eta G_\alpha^{-1} (\theta_\tau) \nabla_\theta L (\theta_\tau), \label{main_eq_update_theta_001_002}
\end{align}
where $\eta$ is a positive number.

\section{Quantum natural gradient} 

To discuss QNG, we explain their quantum counterparts of the Fisher metric, the rescaled R\'enyi divergence, respectively.
Assume that we have $\hat{\rho}_\theta$, which is a quantum state parameterized by $\theta$, and its spectral decomposition is given by $\hat{\rho}_\theta \coloneqq \sum_{i=1}^N p_i | \psi_i \rangle \langle \psi_i |$.
We first introduce the quantum extension of the Fisher metric, Eq.~\eqref{main_eq_def_Fisher_metric_001_001}~\cite{Petz_001, Petz_002, Petz_003, Sagawa_001}.
To this end, we define the following linear operator $\Delta_{\hat{\rho}_\theta}$ as follows: $\Delta_{\hat{\rho}_\theta} \hat{A} \coloneqq \hat{\rho}_\theta \hat{A} \hat{\rho}_\theta^{-1}$.
And we also introduce the nonlinear transformation of $\Delta_{\hat{\rho}_\theta} \hat{A}$:
\begin{align}
	f^{-1} (\Delta_{\hat{\rho}_\theta}) \hat{A} & \coloneqq \sum_{i, j = 1}^N \frac{\langle \psi_i | \hat{A} | \psi_j \rangle}{f (p_i / p_j)} | \psi_i \rangle \langle \psi_j |. \label{main_eq_def_f_inverse_Delta_001_001}
\end{align}
Using Eq.~\eqref{main_eq_def_f_inverse_Delta_001_001}, the quantum counterparts of Eq.~\eqref{main_eq_classical_X_em_001_001} at $\hat{\rho}_\theta$ are defined as follows:
\begin{subequations}
	\begin{align}
		\hat{X}_{\hat{\rho}_\theta}^\mathrm{m}            & \coloneqq X \hat{\rho}_\theta, \label{main_eq_quantum_m_representation_001_001}                                                                \\
		\hat{X}_{\hat{\rho}_\theta, f (\cdot)}^\mathrm{e} & \coloneqq f^{-1} (\Delta_{\hat{\rho}_\theta}) ([X \hat{\rho}_\theta] \hat{\rho}_\theta^{-1}), \label{main_eq_quantum_e_representation_001_001}
	\end{align}
\end{subequations}
where $f (\cdot): \mathbb{R}_{> 0} \to \mathbb{R}_{> 0}$ is a operator-monotone function such that
\begin{subequations}
	\begin{align}
		f (1) & = 1, \label{main_eq_condition_Petz_function_001_011}            \\
		f (t) & = t f (t^{-1}). \label{main_eq_condition_Petz_function_001_012}
	\end{align}
\end{subequations}
Note that $f (\cdot)$ in Eq.~\eqref{main_eq_quantum_e_representation_001_001} is called the Petz function.
Then the quantum Fisher metric is given by
\begin{align}
	g_{\hat{\rho}_\theta, f (\cdot)} (X, Y) & \coloneqq \mathrm{Tr} [\hat{X}_{\hat{\rho}_\theta}^\mathrm{m} \hat{Y}_{\hat{\rho}_\theta, f (\cdot)}^\mathrm{e}].  \label{main_eq_quantum_Fisher_metric_001_001}
\end{align}
The key point of the quantum Fisher metric, Eq.~\eqref{main_eq_quantum_Fisher_metric_001_001}, is its dependence on the operator-monotone function $f (\cdot)$.
Equation~\eqref{main_eq_condition_Petz_function_001_011} implies the classical limit of the quantum Fisher metric goes to the classical Fisher metric without depending on the choice of $f (\cdot)$ and Eq.~\eqref{main_eq_condition_Petz_function_001_012} guarantee for the quantum Fisher metric to be real.

We then consider a quantum extension of Eq.~\eqref{main_eq_def_classical_rescaled_Renyi_divergence_001_001}.
Quantum rescaled sandwiched R\'enyi divergence is defined as follows~\cite{Takahashi_001}:
\begin{align}
	D_\alpha ( \hat{\rho}_{\bar{\theta}} \| \hat{\rho}_\theta ) & \coloneqq \frac{1}{\alpha (\alpha - 1)} \ln \mathrm{Tr} \Big[ \Big( \hat{\rho}_\theta^\frac{1 - \alpha}{2 \alpha} \hat{\rho}_{\bar{\theta}} \hat{\rho}_\theta^\frac{1 - \alpha}{2 \alpha} \Big)^\alpha \Big]. \label{main_eq_def_quantum_rescaled_sandwiched_Renyi_divergence_001_001}
\end{align}
Note that ``sandwiched" means $\hat{\rho}_{\bar{\theta}}$ is sandwiched by $\hat{\rho}_\theta$.
In Ref.~\cite{Amari_002}, quantum $\alpha$-divergence, in which $\hat{\rho}_{\bar{\theta}}$ is not sandwiched by $\hat{\rho}_\theta$, is investigated.
Rescaled sandwiched R\'enyi divergence, Eq.~\eqref{main_eq_def_quantum_rescaled_sandwiched_Renyi_divergence_001_001}, becomes quantum KL divergence~\cite{Umegaki_001} in the limit $\alpha \to 1$: $\lim_{\alpha \to 1} D_\alpha ( \hat{\rho}_{\bar{\theta}} \| \hat{\rho}_\theta ) = D_\mathrm{qKL} ( \hat{\rho}_{\bar{\theta}} \| \hat{\rho}_\theta )$.
Introducing the Petz function given by
\begin{align}
	f_\alpha (t) & \coloneqq (1 - \alpha) \frac{1 - t^\frac{1}{\alpha}}{1 - t^\frac{1 - \alpha}{\alpha}}, \label{main_eq_def_f_alpha_001_001}
\end{align}
Eq.~\eqref{main_eq_quantum_Fisher_metric_001_001} and Eq.~\eqref{main_eq_def_quantum_rescaled_sandwiched_Renyi_divergence_001_001} satisfy the following relation:
\begin{align}
	g_{\hat{\rho}_\theta, f_\alpha (\cdot)} (X, Y) & = \lim_{\bar{\theta} \to \theta} \bar{X} \bar{Y} D_\alpha (\hat{\rho}_{\bar{\theta}} \| \hat{\rho}_\theta), \label{main_eq_quantum_Fisher_metric_derivative_alpha_Renyi_divergence_001_001}
\end{align}
where $\bar{X} \coloneqq \sum_{i=1}^n X^i \bar{\partial}_i$, $\bar{Y} \coloneqq \sum_{i=1}^n Y^i \bar{\partial}_i$.

As the quantum counterpart of Eq.~\eqref{main_optimization_problem_NG_001_001}, let us consider the minimization problem of $L (\theta)$:
\begin{subequations} \label{main_optimization_problem_QNG_001_001}
	\begin{align}
		\theta_{\tau+1}          & = \theta_\tau + \Delta \theta (\epsilon),                                                                                                                                                              \\
		\Delta \theta (\epsilon) & = \argmin_{\Delta \theta: D_\alpha (\hat{\rho}_{\theta_\tau + \Delta \theta} \| \hat{\rho}_{\theta_\tau}) \le \epsilon} L (\theta_\tau + \Delta \theta). \label{main_optimization_problem_QNG_001_012}
	\end{align}
\end{subequations}
In Eq.~\eqref{main_optimization_problem_QNG_001_012}, $\hat{\rho}_\theta$ appears in the constraint instead of $p_\theta (\cdot)$.
By using the lowest-order approximation, Eq.~\eqref{main_optimization_problem_QNG_001_012} can be transformed into
\begin{align}
	\Delta \theta (\epsilon) & \approx \argmin_{\Delta \theta: \frac{1}{2} \Delta \theta^\intercal G_\alpha (\theta_\tau) \Delta \theta \le \epsilon} \nabla_\theta L (\theta_\tau)^\intercal \Delta \theta,
\end{align}
where $G_\alpha (\theta_\tau)$ is the quantum Fisher metric induced by $D_\alpha (\hat{\rho}_{\theta_\tau + \Delta \theta} \| \hat{\rho}_{\theta_\tau})$; that is, it can be rewritten as $[G_\alpha (\theta_\tau)]_{i, j} = g_{\hat{\rho}_\theta, f_\alpha (\cdot)} (\partial_i, \partial_j)$ because of Eq.~\eqref{main_eq_quantum_Fisher_metric_derivative_alpha_Renyi_divergence_001_001}.
To solve Eq.~\eqref{main_optimization_problem_QNG_001_001}, we employ almost the same technique to that used to derive NG.
Then, we reach the same equation in the classical case, Eqs.~\eqref{main_eq_update_theta_001_001} and \eqref{main_eq_update_theta_001_002}, but $G_\alpha^{-1} (\theta_\tau)$ is computed from Eq.~\eqref{main_eq_quantum_Fisher_metric_derivative_alpha_Renyi_divergence_001_001}.
Note that we do not need to limit ourselves to Eq.~\eqref{main_eq_def_quantum_rescaled_sandwiched_Renyi_divergence_001_001} though in this paper we focus on Eq.~\eqref{main_eq_def_quantum_rescaled_sandwiched_Renyi_divergence_001_001}.

\section{Monotonicity} 

Characterizing metrics and divergences, monotonicity is an essential concept~\cite{Petz_001, Hiai_002, Hiai_007}.
For example, the monotonicity of quantum KL divergence is considered to be quite natural since it implies that information is not gained via CPTP maps.

We review some important properties of operator functions~\cite{Bhatia_001, Horn_001, Hiai_002, Hiai_007}.
We first state the definition of the monotonicity of a function: that is, $f (\cdot)$ is called monotone if and only if
\begin{align}
	\hat{A} \succeq \hat{B} & \Rightarrow f (\hat{A}) \succeq f (\hat{B}), \label{main_eq_def_operator_monotone_function_001_001}
\end{align}
where $\hat{A} \succeq \hat{B} :\Leftrightarrow \text{$\forall | \psi \rangle \in \mathcal{H}$, $\langle \psi | (\hat{A} - \hat{B}) | \psi \rangle \ge 0$}$.

Metric $g_{\hat{\rho}} (X, Y)$ is called a monotone metric if and only if it satisfies the following relation for arbitrary CPTP map $\gamma (\cdot)$: $g_{\hat{\rho}} (X, X) \ge g_{\gamma (\hat{\rho})} (\gamma_* (X), \gamma_* (X))$ where $\gamma_* (\cdot)$ is the pushforward associated with $\gamma (\cdot)$.
Note that Eq.~\eqref{main_eq_def_f_alpha_001_001} is operator-monotone if and only if $\alpha \in (- \infty, -1] \cup [\frac{1}{2}, \infty)$ and Eq.~\eqref{main_eq_quantum_Fisher_metric_derivative_alpha_Renyi_divergence_001_001} is a monotone metric~\cite{Takahashi_001}.

We define the following partial order on $f (\cdot): \mathbb{R}_{\ge 0} \to \mathbb{R}_{\ge 0}$:
\begin{align}
	\text{$f (\cdot) \preceq \tilde{f} (\cdot)$} & :\Leftrightarrow \text{$\forall t \in \mathbb{R}_{\ge 0}$, $f (t) \le \tilde{f} (t)$}. \label{main_eq_def_order_operator_monotone_functions_001_001}
\end{align}
For monotone functions, the following theorem holds~\cite{Petz_001}:
\begin{theorem} \label{main_theorem_maximum_minimum_elements_order_operator_monotone_functions_001_001}
	$f_\mathrm{SLD} (\cdot)$ and $f_\mathrm{rRLD} (\cdot)$ are the maximum and the minimum elements with respect to Eq.~\eqref{main_eq_def_order_operator_monotone_functions_001_001} under the condition of monotonicity, Eq.~\eqref{main_eq_def_operator_monotone_function_001_001}, $f (1) = 1$, and $f (t) = t f (t^{-1})$.
	That is, we have the following relation for any $f (t)$ such that the condition of monotonicity, Eq.~\eqref{main_eq_def_operator_monotone_function_001_001}, $f (1) = 1$, and $f (t) = t f (t^{-1})$:
	\begin{align}
		f_\mathrm{rRLD} (\cdot) \preceq f (\cdot) \preceq f_\mathrm{SLD} (\cdot), \label{main_eq_maximum_minimum_elements_order_operator_monotone_functions_001_001}
	\end{align}
	where $f_\mathrm{SLD} (\cdot)$ and $f_\mathrm{rRLD} (\cdot)$ are the Petz functions for the SLD and rRLD metrics defined, respectively, as $f_\mathrm{SLD} (t) \coloneqq \frac{1 + t}{2}$,	$f_\mathrm{rRLD} (t) \coloneqq \frac{2 t}{t + 1}$.
\end{theorem}

We review some important properties of Eq.~\eqref{main_eq_def_f_alpha_001_001}.
For simplicity, we substitute $\alpha = \frac{1}{\beta}$ into Eq.~\eqref{main_eq_def_f_alpha_001_001}:
\begin{align}
	f_\frac{1}{\beta} (t) & = \bigg( 1 - \frac{1}{\beta} \bigg) \frac{1 - t^\beta}{1 - t^{\beta - 1}}. \label{main_eq_def_f_beta_001_001}
\end{align}
Two important theorems on Eq.~\eqref{main_eq_def_f_beta_001_001} are shown in Ref.~\cite{Takahashi_001}.
The first one is as follows~\cite{Takahashi_001}:
\begin{theorem} \label{main_theorem_monotonicity_Petz_function_quantum_rescaled_sandwiched_Renyi_divergence_001_001}
	Eq.~\eqref{main_eq_def_f_beta_001_001} monotonically increases when $\beta$ increased.
\end{theorem}
Thm.~\ref{main_theorem_monotonicity_Petz_function_quantum_rescaled_sandwiched_Renyi_divergence_001_001} implies that $f_\alpha (\cdot)$, Eq.~\eqref{main_eq_def_f_alpha_001_001}, is monotonically decreasing with $\alpha$ increasing, but $\alpha = 0$ is singular.
Thus, $f_\alpha (\cdot)$, Eq.~\eqref{main_eq_def_f_alpha_001_001}, monotonically increases toward $\alpha \to 0+$.

The second one is as follows~\cite{Takahashi_001}:
\begin{theorem} \label{main_theorem_monotonicity_Petz_function_quantum_rescaled_sandwiched_Renyi_divergence_002_001}
	Eq.~\eqref{main_eq_def_f_beta_001_001} is operator monotone if and only if $\beta \in [-1, 2]$.
\end{theorem}
Thm.~\ref{main_theorem_monotonicity_Petz_function_quantum_rescaled_sandwiched_Renyi_divergence_002_001} means that $f_\alpha (\cdot)$, Eq.~\eqref{main_eq_def_f_alpha_001_001}, is operator monotone for $\alpha \in (\infty, -1] \cup [1/2, \infty)$.

\begin{figure}[t]
	\centering
	\includegraphics[scale=0.60]{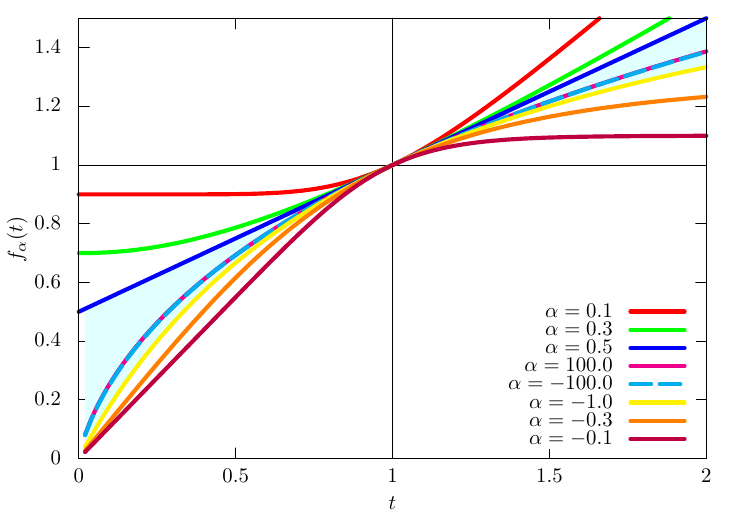}
	\caption{$f_\alpha (t)$, Eq.~\eqref{main_eq_def_f_alpha_001_001}, with $\alpha = 0.1, 0.3, 0.5, 100.0, -100.0, -1.0, -0.3, -0.1$. Note that $\alpha = 0.5$ and $\alpha = -1.0$ yield the SLD and rRLD metrics, respectively. The regime of the monotone Petz functions is highlighted by light cyan.}
	\label{main_fig_gnuplot_operator_monotone_functions_002_001}
\end{figure}

In Fig.~\ref{main_fig_gnuplot_operator_monotone_functions_002_001}, we show $f_\alpha (t)$, Eq.~\eqref{main_eq_def_f_alpha_001_001}, for several different $\alpha$ and highlight the regime of the monotone Petz functions by light cyan.
Note that $f_\alpha (t)$, Eq.~\eqref{main_eq_def_f_alpha_001_001}, with $\alpha = 1/2, -1$ are the Petz functions for the SLD and rRLD metrics, respectively.
Furthermore, we also have $\lim_{\alpha \to \pm \infty} f_\alpha (t) = \frac{t \ln t}{t - 1}$~\cite{Takahashi_001}.

\section{Fisher metrics and speed of QNG} 

We here explain the relation between the quantum Fisher metrics and the speed of QNG.
In general, the following theorem holds:
\begin{theorem} \label{main_theorem_relation_Petz_function_quantum_Fisher_metrics_001_001}
	Let us assume that $f (\cdot)$ and $g_{\hat{\rho}_\theta, f (\cdot)} (\cdot, \cdot)$ are related via Eq.~\eqref{main_eq_quantum_Fisher_metric_001_001}.
	Then, we have
	\begin{align}
		f (\cdot) \preceq \tilde{f} (\cdot) & \Leftrightarrow g_{\hat{\rho}_\theta, f (\cdot)} (\cdot, \cdot) \succeq g_{\hat{\rho}_\theta, \tilde{f} (\cdot)} (\cdot, \cdot), \label{main_eq_relation_Petz_function_quantum_Fisher_metric_001_001}
	\end{align}
	where $\text{$g (\cdot, \cdot) \succeq \tilde{g} (\cdot, \cdot)$} :\Leftrightarrow \text{$\forall X$, $g (X, X) \ge \tilde{g} (X, X)$}$.
\end{theorem}
The proof of Thm.~\ref{main_theorem_relation_Petz_function_quantum_Fisher_metrics_001_001} is shown in Sec.~\ref{main_appendix_proof_001_001} in Appendix~\ref{main_appendix_proof_001_001}.

We define the following diagonal metric:
\begin{align}
	\bar{g}_{\hat{\rho}_\theta, f (\cdot)} (\partial_i, \partial_j) & \coloneqq
	\begin{cases}
		g_{\hat{\rho}_\theta, f (\cdot)} (\partial_i, \partial_j) & (i = j),   \\
		0                                                         & (i \ne j).
	\end{cases} \label{main_eq_def_tilde_g_001_001}
\end{align}
In partial usages, Eq.~\eqref{main_eq_def_tilde_g_001_001} is of great importance and often used instead of Eq.~\eqref{main_eq_quantum_Fisher_metric_001_001} because the computation of the inverse matrix of Eq.~\eqref{main_eq_quantum_Fisher_metric_001_001} is numerically costly.
Similarly to Thm.~\ref{main_theorem_relation_Petz_function_quantum_Fisher_metrics_001_001}, the following theorem holds:
\begin{theorem} \label{main_theorem_relation_Petz_function_quantum_Fisher_metrics_002_001}
	The following relation holds for Eq.~\eqref{main_eq_def_tilde_g_001_001}:
	\begin{align}
		f (t) \preceq \tilde{f} (\cdot) & \Leftrightarrow \bar{g}_{\hat{\rho}_\theta, f (\cdot)} (\cdot, \cdot) \succeq \bar{g}_{\hat{\rho}_\theta, \tilde{f} (\cdot)} (\cdot, \cdot). \label{main_eq_relation_Petz_function_quantum_Fisher_metric_002_001}
	\end{align}
\end{theorem}
The proof of Thm.~\ref{main_theorem_relation_Petz_function_quantum_Fisher_metrics_002_001} is almost the same with Thm.~\ref{main_theorem_relation_Petz_function_quantum_Fisher_metrics_001_001}.

We then discuss the dependence of the speed of QNG on the quantum Fisher metrics.
Metrics $g (\cdot, \cdot)$ and $\tilde{g} (\cdot, \cdot)$ satisfy the following relation: $g (\cdot, \cdot) \succeq \tilde{g} (\cdot, \cdot) \Leftrightarrow g^{-1} (\cdot, \cdot) \preceq \tilde{g}^{-1} (\cdot, \cdot)$.
In the case of Eq.~\eqref{main_eq_update_theta_001_001}, we have
\begin{align}
	L (\theta_{\tau+1}) - L (\theta_\tau) & = - \sqrt{2 \epsilon \nabla_\theta L (\theta_\tau)^\intercal G_\alpha^{-1} (\theta_\tau) \nabla_\theta L (\theta_\tau)}.
\end{align}
And in the case of Eq.~\eqref{main_eq_update_theta_001_002}, we also have
\begin{align}
	L (\theta_{\tau+1}) - L (\theta_\tau) & = - \eta \nabla_\theta L (\theta_\tau)^\intercal G_\alpha^{-1} (\theta_\tau) \nabla_\theta L (\theta_\tau).
\end{align}
Thus, for both of Eqs.~\eqref{main_eq_update_theta_001_001} and \eqref{main_eq_update_theta_001_002}, $G_\alpha (\theta_\tau)$ decrease the value of $L (\theta_\tau)$ faster than $G_{\alpha'} (\theta_\tau)$ when $G_\alpha (\theta_\tau) \preceq G_{\alpha'} (\theta_\tau)$.

\section{Numerical simulations} 

We show numerical simulations to support the main claim of this paper.
We begin with the setup of numerical simulations.
Using $\theta \coloneqq [\theta_1, \theta_2, \theta_3]^\intercal$, we define the following parameterized quantum state: $\hat{\rho}_\theta \coloneqq \hat{U} (\theta) \hat{\rho}_\mathrm{ini} \hat{U}^\dagger (\theta)$ where $\hat{U} (\theta) \coloneqq \hat{R}_z (\theta_3) \hat{R}_y (\theta_2) \hat{R}_z (\theta_1)$, $\hat{R}_z (\phi) \coloneqq \exp ( - i \phi \hat{\sigma}_z / 2 )$, and $\hat{R}_y (\phi) \coloneqq \exp ( - i \phi \hat{\sigma}_y / 2 )$.
We also define $\hat{\rho}_\mathrm{ini} \coloneqq \frac{1}{2} (\hat{1} + x \hat{\sigma}_x + y \hat{\sigma}_y + z \hat{\sigma}_z)$ where $\hat{1}$ is the $2 \times 2$ identity operator and $x^2 + y^2 + z^2 < 1$.
Then, we consider the following minimization problem with respect to $\theta$:
\begin{align}
	\min_\theta L (\theta),
\end{align}
where $L (\theta) \coloneqq \mathrm{Tr} [(\hat{\rho}_\theta - \hat{\rho}_{\theta_*})^\dagger (\hat{\rho}_\theta - \hat{\rho}_{\theta_*})]$.
To stabilize the numerical calculations, we conduct the following operations to the density operator and metric: $\hat{\rho}_\theta \leftarrow (1 - \delta) \hat{\rho}_\theta + \delta \frac{\hat{1}}{N}$, and $g_{\hat{\rho}_\theta, f (\cdot)} (\cdot, \cdot) \leftarrow (1 - \xi) g_{\hat{\rho}_\theta, f (\cdot)} (\cdot, \cdot) + \xi I$ where $I$ is the identity matrix whose size is identical to that of $g$, and $\delta$ and $\eta$ are tiny positive numbers.

\begin{figure}[t]
	\centering
	\includegraphics[scale=0.60]{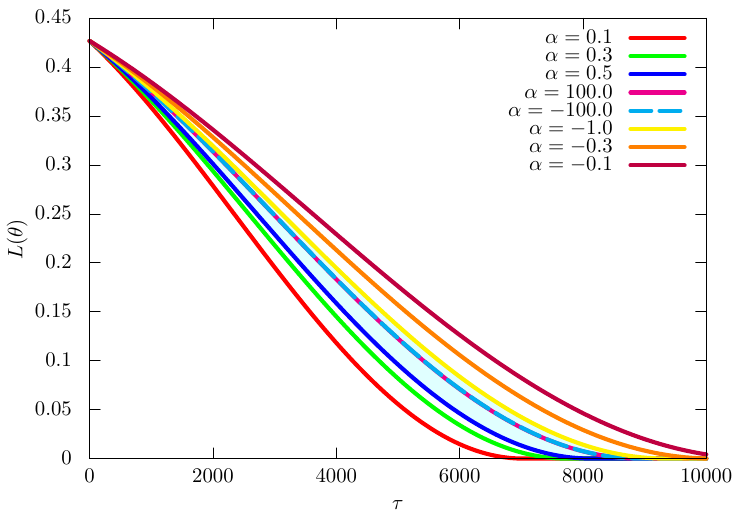}
	\caption{Cost functions for several $\alpha$ in the case of Eq.~\eqref{main_eq_update_theta_001_001}. We set $\epsilon = 1.0 \times 10^{-8}$, $\xi = 1.0 \times 10^{-3}$, and $\delta = 1.0 \times 10^{-3}$. The regime of the monotone metrics is highlighted by light cyan.}
	\label{main_fig_performance_QNG_rot-states_001_001}
\end{figure}

\begin{figure}[t]
	\centering
	\includegraphics[scale=0.60]{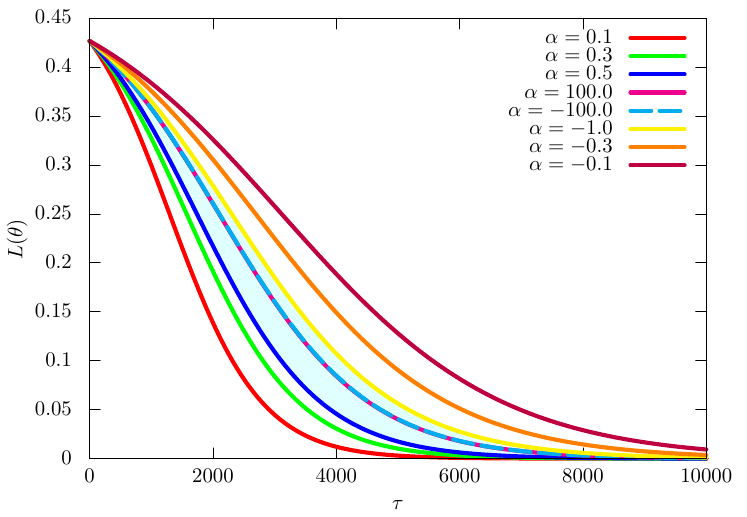}
	\caption{Cost functions for several $\alpha$ in the case of Eq.~\eqref{main_eq_update_theta_001_002}. We set $\eta = 5.0 \times 10^{-4}$, $\xi = 1.0 \times 10^{-3}$, and $\delta = 1.0 \times 10^{-3}$. The regime of the monotone metrics is highlighted by light cyan.}
	\label{main_fig_performance_QNG_rot-states_002_001}
\end{figure}

In the following calculations, we set $[x, y, z] = [0.5, 0.0, 0.0]$, $\theta_0 = [\pi / 2, \pi / 2, \pi / 4]^\intercal$, and $\theta_* = [0.0, 0.0, 0.0]^\intercal$.
In Fig.~\ref{main_fig_performance_QNG_rot-states_001_001}, we plot the time evolution of the cost function in the case of Eq.~\eqref{main_eq_update_theta_001_001} and highlight the regime of the monotone Petz functions by light cyan.
In Fig.~\ref{main_fig_performance_QNG_rot-states_002_001}, we show the time evolution of the cost function in the case of Eq.~\eqref{main_eq_update_theta_001_002}.
In both cases, QNG with $\alpha \in (0.0, 0.5)$ outperforms SLD-based QNG proposed in Refs.~\cite{Stokes_001, Koczor_001}.
Furthermore, these figures show that the gaps between different values of $\alpha$ becomes large in the case of Eq.~\eqref{main_eq_update_theta_001_002} than Eq.~\eqref{main_eq_update_theta_001_001}.
However, this phenomenon depends on the maximum eigenvalue of the quantum Fisher metric; thus, we have to choose either of Eq.~\eqref{main_eq_update_theta_001_001} or Eq.~\eqref{main_eq_update_theta_001_002} depending on problems.

\section{Conclusions}

In this paper, we have presented a generalization of QNG based on QIG.
Notably, the existing literature on QNG relies on the SLD metric, which is not the exclusive definition of the logarithmic derivative of density operators.
In QNG, the e-representation, specified by the Petz function, aligns with the logarithmic derivative of density operators, and it identifies the quantum Fisher metric.
In our work, we utilize the non-monotone quantum Fisher metrics to generalize QNG, successfully enhancing the speed of QNG.
Monotonicity has been believed to be a necessary condition to construct a geometry in physics; however, we have clarified that it is not necessary; in fact, it should be eliminated for optimization purposes.
Our findings have shed light on the fact that we need to be cautious even with mathematical principles to advance science.

\begin{acknowledgments}
	H.M. was supported by JSPS KAKENHI Grant Number JP23H04489.
	H.M. thanks Akio Fujiwara, Lei Wang, Yusuke Nomura, and Shunsuke Daimon for fruitful discussions.
\end{acknowledgments}

\appendix
\renewcommand{\theequation}{\Alph{section}.\arabic{equation}}

\section{Proof of Thm.~\ref{main_theorem_relation_Petz_function_quantum_Fisher_metrics_001_001}} \label{main_appendix_proof_001_001}

The proof of Thm.~\ref{main_theorem_relation_Petz_function_quantum_Fisher_metrics_001_001} is as follows:
\begin{proof}
	We show Eq.~\eqref{main_eq_relation_Petz_function_quantum_Fisher_metric_001_001}.
	From Eq.~\eqref{main_eq_quantum_Fisher_metric_001_001}, we have
	\begin{align}
		g_{\hat{\rho}_\theta, f (\cdot)} (X, X) & = \sum_{i, j = 1}^N \frac{1}{p_j f (p_i / p_j)} \langle \psi_j | \hat{X}_{\hat{\rho}_\theta}^\mathrm{m} | \psi_i \rangle \langle \psi_i | \hat{X}_{\hat{\rho}_\theta}^\mathrm{m} | \psi_j \rangle \\
		                                        & = \sum_{i, j = 1}^N \frac{1}{p_j f (p_i / p_j)} | \langle \psi_j | \hat{X}_{\hat{\rho}_\theta}^\mathrm{m} | \psi_i \rangle |^2.
	\end{align}
	Since $f (\cdot) \preceq \tilde{f} (\cdot)$, we have
	\begin{align}
		g_{\hat{\rho}_\theta, f (\cdot)} (X, X) & = \sum_{i, j = 1}^N \frac{1}{p_j f (p_i / p_j)} | \langle \psi_j | \hat{X}_{\hat{\rho}_\theta}^\mathrm{m} | \psi_i \rangle |^2           \\
		                                        & \ge \sum_{i, j = 1}^N \frac{1}{p_j \tilde{f} (p_i / p_j)} | \langle \psi_j | \hat{X}_{\hat{\rho}_\theta}^\mathrm{m} | \psi_i \rangle |^2 \\
		                                        & = g_{\hat{\rho}_\theta, \tilde{f} (\cdot)} (X, X).
	\end{align}
	Thus, Eq.~\eqref{main_eq_relation_Petz_function_quantum_Fisher_metric_001_001} is proved.
\end{proof}

\bibliography{paper_QNG_without_monotonicity_999_001}

\begin{thebibliography}{37}
\expandafter\ifx\csname natexlab\endcsname\relax\def\natexlab#1{#1}\fi
\expandafter\ifx\csname bibnamefont\endcsname\relax
  \def\bibnamefont#1{#1}\fi
\expandafter\ifx\csname bibfnamefont\endcsname\relax
  \def\bibfnamefont#1{#1}\fi
\expandafter\ifx\csname citenamefont\endcsname\relax
  \def\citenamefont#1{#1}\fi
\expandafter\ifx\csname url\endcsname\relax
  \def\url#1{\texttt{#1}}\fi
\expandafter\ifx\csname urlprefix\endcsname\relax\def\urlprefix{URL }\fi
\providecommand{\bibinfo}[2]{#2}
\providecommand{\eprint}[2][]{\url{#2}}

\bibitem[{\citenamefont{Nocedal and Wright}(2009)}]{Nocedal_001}
\bibinfo{author}{\bibfnamefont{J.}~\bibnamefont{Nocedal}} \bibnamefont{and}
  \bibinfo{author}{\bibfnamefont{S.}~\bibnamefont{Wright}},
  \emph{\bibinfo{title}{Numerical Optimization}}, Springer Series in Operations
  Research and Financial Engineering (\bibinfo{publisher}{Springer New York},
  \bibinfo{year}{2009}), ISBN \bibinfo{isbn}{9781493937110},
  \urlprefix\url{https://books.google.co.jp/books?id=en-YswEACAAJ}.

\bibitem[{\citenamefont{Amari and Nagaoka}(2000)}]{Amari_002}
\bibinfo{author}{\bibfnamefont{S.-i.} \bibnamefont{Amari}} \bibnamefont{and}
  \bibinfo{author}{\bibfnamefont{H.}~\bibnamefont{Nagaoka}},
  \emph{\bibinfo{title}{Methods of information geometry}}, vol.
  \bibinfo{volume}{191} (\bibinfo{publisher}{American Mathematical Soc.},
  \bibinfo{year}{2000}).

\bibitem[{\citenamefont{Amari}(2016)}]{Amari_004}
\bibinfo{author}{\bibfnamefont{S.-i.} \bibnamefont{Amari}},
  \emph{\bibinfo{title}{Information Geometry and Its Applications}}, Applied
  Mathematical Sciences (\bibinfo{publisher}{Springer Japan},
  \bibinfo{year}{2016}), ISBN \bibinfo{isbn}{9784431559788},
  \urlprefix\url{https://books.google.co.jp/books?id=UkSFCwAAQBAJ}.

\bibitem[{\citenamefont{Amari}(1998)}]{Amari_005}
\bibinfo{author}{\bibfnamefont{S.-I.} \bibnamefont{Amari}},
  \bibinfo{journal}{Neural computation} \textbf{\bibinfo{volume}{10}},
  \bibinfo{pages}{251} (\bibinfo{year}{1998}).

\bibitem[{\citenamefont{Bishop}(2006)}]{Bishop_001}
\bibinfo{author}{\bibfnamefont{C.~M.} \bibnamefont{Bishop}},
  \emph{\bibinfo{title}{Pattern recognition and machine learning}}
  (\bibinfo{publisher}{springer}, \bibinfo{year}{2006}).

\bibitem[{\citenamefont{Murphy}(2022)}]{Murphy_002}
\bibinfo{author}{\bibfnamefont{K.}~\bibnamefont{Murphy}},
  \emph{\bibinfo{title}{Probabilistic Machine Learning: An Introduction}},
  Adaptive Computation and Machine Learning series (\bibinfo{publisher}{MIT
  Press}, \bibinfo{year}{2022}), ISBN \bibinfo{isbn}{9780262369305},
  \urlprefix\url{https://books.google.co.jp/books?id=OyYuEAAAQBAJ}.

\bibitem[{\citenamefont{Murphy}(2023)}]{Murphy_003}
\bibinfo{author}{\bibfnamefont{K.}~\bibnamefont{Murphy}},
  \emph{\bibinfo{title}{Probabilistic Machine Learning: Advanced Topics}},
  Adaptive Computation and Machine Learning series (\bibinfo{publisher}{MIT
  Press}, \bibinfo{year}{2023}), ISBN \bibinfo{isbn}{9780262048439},
  \urlprefix\url{https://books.google.co.jp/books?id=8iycEAAAQBAJ}.

\bibitem[{\citenamefont{Martens}(2020)}]{Martens_001}
\bibinfo{author}{\bibfnamefont{J.}~\bibnamefont{Martens}},
  \bibinfo{journal}{The Journal of Machine Learning Research}
  \textbf{\bibinfo{volume}{21}}, \bibinfo{pages}{5776} (\bibinfo{year}{2020}).

\bibitem[{\citenamefont{Preskill}(2018)}]{Preskill_001}
\bibinfo{author}{\bibfnamefont{J.}~\bibnamefont{Preskill}},
  \bibinfo{journal}{Quantum} \textbf{\bibinfo{volume}{2}}, \bibinfo{pages}{79}
  (\bibinfo{year}{2018}).

\bibitem[{\citenamefont{Mitarai et~al.}(2018)\citenamefont{Mitarai, Negoro,
  Kitagawa, and Fujii}}]{Mitarai_001}
\bibinfo{author}{\bibfnamefont{K.}~\bibnamefont{Mitarai}},
  \bibinfo{author}{\bibfnamefont{M.}~\bibnamefont{Negoro}},
  \bibinfo{author}{\bibfnamefont{M.}~\bibnamefont{Kitagawa}}, \bibnamefont{and}
  \bibinfo{author}{\bibfnamefont{K.}~\bibnamefont{Fujii}},
  \bibinfo{journal}{Phys. Rev. A} \textbf{\bibinfo{volume}{98}},
  \bibinfo{pages}{032309} (\bibinfo{year}{2018}),
  \urlprefix\url{https://link.aps.org/doi/10.1103/PhysRevA.98.032309}.

\bibitem[{\citenamefont{Schuld et~al.}(2020)\citenamefont{Schuld, Bocharov,
  Svore, and Wiebe}}]{Schuld_001}
\bibinfo{author}{\bibfnamefont{M.}~\bibnamefont{Schuld}},
  \bibinfo{author}{\bibfnamefont{A.}~\bibnamefont{Bocharov}},
  \bibinfo{author}{\bibfnamefont{K.~M.} \bibnamefont{Svore}}, \bibnamefont{and}
  \bibinfo{author}{\bibfnamefont{N.}~\bibnamefont{Wiebe}},
  \bibinfo{journal}{Phys. Rev. A} \textbf{\bibinfo{volume}{101}},
  \bibinfo{pages}{032308} (\bibinfo{year}{2020}),
  \urlprefix\url{https://link.aps.org/doi/10.1103/PhysRevA.101.032308}.

\bibitem[{\citenamefont{Miyahara and Roychowdhury}(2022)}]{Miyahara_004}
\bibinfo{author}{\bibfnamefont{H.}~\bibnamefont{Miyahara}} \bibnamefont{and}
  \bibinfo{author}{\bibfnamefont{V.}~\bibnamefont{Roychowdhury}},
  \bibinfo{journal}{Scientific Reports} \textbf{\bibinfo{volume}{12}},
  \bibinfo{pages}{19520} (\bibinfo{year}{2022}).

\bibitem[{\citenamefont{Scherer and Scherer}(2017)}]{Scherer_001}
\bibinfo{author}{\bibfnamefont{P.~O.} \bibnamefont{Scherer}} \bibnamefont{and}
  \bibinfo{author}{\bibfnamefont{P.~O.} \bibnamefont{Scherer}},
  \emph{\bibinfo{title}{Computational physics: simulation of classical and
  quantum systems}} (\bibinfo{publisher}{Springer}, \bibinfo{year}{2017}).

\bibitem[{\citenamefont{Or{\'u}s}(2019)}]{Orus_001}
\bibinfo{author}{\bibfnamefont{R.}~\bibnamefont{Or{\'u}s}},
  \bibinfo{journal}{Nature Reviews Physics} \textbf{\bibinfo{volume}{1}},
  \bibinfo{pages}{538} (\bibinfo{year}{2019}).

\bibitem[{\citenamefont{Stokes et~al.}(2020)\citenamefont{Stokes, Izaac,
  Killoran, and Carleo}}]{Stokes_001}
\bibinfo{author}{\bibfnamefont{J.}~\bibnamefont{Stokes}},
  \bibinfo{author}{\bibfnamefont{J.}~\bibnamefont{Izaac}},
  \bibinfo{author}{\bibfnamefont{N.}~\bibnamefont{Killoran}}, \bibnamefont{and}
  \bibinfo{author}{\bibfnamefont{G.}~\bibnamefont{Carleo}},
  \bibinfo{journal}{Quantum} \textbf{\bibinfo{volume}{4}}, \bibinfo{pages}{269}
  (\bibinfo{year}{2020}).

\bibitem[{\citenamefont{Koczor and Benjamin}(2022)}]{Koczor_001}
\bibinfo{author}{\bibfnamefont{B.}~\bibnamefont{Koczor}} \bibnamefont{and}
  \bibinfo{author}{\bibfnamefont{S.~C.} \bibnamefont{Benjamin}},
  \bibinfo{journal}{Phys. Rev. A} \textbf{\bibinfo{volume}{106}},
  \bibinfo{pages}{062416} (\bibinfo{year}{2022}),
  \urlprefix\url{https://link.aps.org/doi/10.1103/PhysRevA.106.062416}.

\bibitem[{\citenamefont{Sorella and Capriotti}(2000)}]{Sorella_001}
\bibinfo{author}{\bibfnamefont{S.}~\bibnamefont{Sorella}} \bibnamefont{and}
  \bibinfo{author}{\bibfnamefont{L.}~\bibnamefont{Capriotti}},
  \bibinfo{journal}{Phys. Rev. B} \textbf{\bibinfo{volume}{61}},
  \bibinfo{pages}{2599} (\bibinfo{year}{2000}),
  \urlprefix\url{https://link.aps.org/doi/10.1103/PhysRevB.61.2599}.

\bibitem[{\citenamefont{Sorella}(2001)}]{Sorella_002}
\bibinfo{author}{\bibfnamefont{S.}~\bibnamefont{Sorella}},
  \bibinfo{journal}{Phys. Rev. B} \textbf{\bibinfo{volume}{64}},
  \bibinfo{pages}{024512} (\bibinfo{year}{2001}),
  \urlprefix\url{https://link.aps.org/doi/10.1103/PhysRevB.64.024512}.

\bibitem[{\citenamefont{Mazzola et~al.}(2012)\citenamefont{Mazzola, Zen, and
  Sorella}}]{Mazzola_001}
\bibinfo{author}{\bibfnamefont{G.}~\bibnamefont{Mazzola}},
  \bibinfo{author}{\bibfnamefont{A.}~\bibnamefont{Zen}}, \bibnamefont{and}
  \bibinfo{author}{\bibfnamefont{S.}~\bibnamefont{Sorella}},
  \bibinfo{journal}{The Journal of Chemical Physics}
  \textbf{\bibinfo{volume}{137}}, \bibinfo{pages}{134112}
  (\bibinfo{year}{2012}), ISSN \bibinfo{issn}{0021-9606},
  \eprint{https://pubs.aip.org/aip/jcp/article-pdf/doi/10.1063/1.4755992/15454421/134112\_1\_online.pdf},
  \urlprefix\url{https://doi.org/10.1063/1.4755992}.

\bibitem[{\citenamefont{Park and Kastoryano}(2020)}]{Park_001}
\bibinfo{author}{\bibfnamefont{C.-Y.} \bibnamefont{Park}} \bibnamefont{and}
  \bibinfo{author}{\bibfnamefont{M.~J.} \bibnamefont{Kastoryano}},
  \bibinfo{journal}{Phys. Rev. Res.} \textbf{\bibinfo{volume}{2}},
  \bibinfo{pages}{023232} (\bibinfo{year}{2020}),
  \urlprefix\url{https://link.aps.org/doi/10.1103/PhysRevResearch.2.023232}.

\bibitem[{\citenamefont{Xie et~al.}(2023)\citenamefont{Xie, Zhang, and
  Wang}}]{Xie_001}
\bibinfo{author}{\bibfnamefont{H.}~\bibnamefont{Xie}},
  \bibinfo{author}{\bibfnamefont{L.}~\bibnamefont{Zhang}}, \bibnamefont{and}
  \bibinfo{author}{\bibfnamefont{L.}~\bibnamefont{Wang}},
  \bibinfo{journal}{SciPost Phys.} \textbf{\bibinfo{volume}{14}},
  \bibinfo{pages}{154} (\bibinfo{year}{2023}),
  \urlprefix\url{https://scipost.org/10.21468/SciPostPhys.14.6.154}.

\bibitem[{\citenamefont{Lopatnikova et~al.}(2021)\citenamefont{Lopatnikova,
  Tran, and Sisson}}]{Lopatnikova_001}
\bibinfo{author}{\bibfnamefont{A.}~\bibnamefont{Lopatnikova}},
  \bibinfo{author}{\bibfnamefont{M.-N.} \bibnamefont{Tran}}, \bibnamefont{and}
  \bibinfo{author}{\bibfnamefont{S.~A.} \bibnamefont{Sisson}},
  \bibinfo{journal}{arXiv preprint arXiv:2112.06587}  (\bibinfo{year}{2021}).

\bibitem[{\citenamefont{Miyahara et~al.}(2020)\citenamefont{Miyahara, Aihara,
  and Lechner}}]{Miyahara_001}
\bibinfo{author}{\bibfnamefont{H.}~\bibnamefont{Miyahara}},
  \bibinfo{author}{\bibfnamefont{K.}~\bibnamefont{Aihara}}, \bibnamefont{and}
  \bibinfo{author}{\bibfnamefont{W.}~\bibnamefont{Lechner}},
  \bibinfo{journal}{Physical Review A} \textbf{\bibinfo{volume}{101}},
  \bibinfo{pages}{012326} (\bibinfo{year}{2020}).

\bibitem[{\citenamefont{Miyahara and Sughiyama}(2018)}]{Miyahara_002}
\bibinfo{author}{\bibfnamefont{H.}~\bibnamefont{Miyahara}} \bibnamefont{and}
  \bibinfo{author}{\bibfnamefont{Y.}~\bibnamefont{Sughiyama}},
  \bibinfo{journal}{Physical Review A} \textbf{\bibinfo{volume}{98}},
  \bibinfo{pages}{022330} (\bibinfo{year}{2018}).

\bibitem[{\citenamefont{Miyahara and Roychowdhury}(2023)}]{Miyahara_003}
\bibinfo{author}{\bibfnamefont{H.}~\bibnamefont{Miyahara}} \bibnamefont{and}
  \bibinfo{author}{\bibfnamefont{V.}~\bibnamefont{Roychowdhury}},
  \bibinfo{journal}{Proceedings of the National Academy of Sciences}
  \textbf{\bibinfo{volume}{120}}, \bibinfo{pages}{e2212660120}
  (\bibinfo{year}{2023}).

\bibitem[{\citenamefont{Van~Erven and Harremos}(2014)}]{Erven_001}
\bibinfo{author}{\bibfnamefont{T.}~\bibnamefont{Van~Erven}} \bibnamefont{and}
  \bibinfo{author}{\bibfnamefont{P.}~\bibnamefont{Harremos}},
  \bibinfo{journal}{IEEE Transactions on Information Theory}
  \textbf{\bibinfo{volume}{60}}, \bibinfo{pages}{3797} (\bibinfo{year}{2014}).

\bibitem[{\citenamefont{Mizohata et~al.}(2023)\citenamefont{Mizohata,
  Kobayashi, Bouchard, and Miyahara}}]{Mizohata_001}
\bibinfo{author}{\bibfnamefont{T.}~\bibnamefont{Mizohata}},
  \bibinfo{author}{\bibfnamefont{T.~J.} \bibnamefont{Kobayashi}},
  \bibinfo{author}{\bibfnamefont{L.-S.} \bibnamefont{Bouchard}},
  \bibnamefont{and} \bibinfo{author}{\bibfnamefont{H.}~\bibnamefont{Miyahara}},
  \bibinfo{journal}{arXiv preprint arXiv:2309.10334}  (\bibinfo{year}{2023}).

\bibitem[{\citenamefont{Petz}(1996)}]{Petz_001}
\bibinfo{author}{\bibfnamefont{D.}~\bibnamefont{Petz}},
  \bibinfo{journal}{Linear Algebra and its Applications}
  \textbf{\bibinfo{volume}{244}}, \bibinfo{pages}{81} (\bibinfo{year}{1996}),
  ISSN \bibinfo{issn}{0024-3795},
  \urlprefix\url{https://www.sciencedirect.com/science/article/pii/0024379594002118}.

\bibitem[{\citenamefont{Petz}(2007)}]{Petz_002}
\bibinfo{author}{\bibfnamefont{D.}~\bibnamefont{Petz}},
  \emph{\bibinfo{title}{Quantum information theory and quantum statistics}}
  (\bibinfo{publisher}{Springer Science \& Business Media},
  \bibinfo{year}{2007}).

\bibitem[{\citenamefont{Petz and Ghinea}(2011)}]{Petz_003}
\bibinfo{author}{\bibfnamefont{D.}~\bibnamefont{Petz}} \bibnamefont{and}
  \bibinfo{author}{\bibfnamefont{C.}~\bibnamefont{Ghinea}}, in
  \emph{\bibinfo{booktitle}{Quantum probability and related topics}}
  (\bibinfo{publisher}{World Scientific}, \bibinfo{year}{2011}), pp.
  \bibinfo{pages}{261--281}.

\bibitem[{\citenamefont{Sagawa}(2022)}]{Sagawa_001}
\bibinfo{author}{\bibfnamefont{T.}~\bibnamefont{Sagawa}},
  \emph{\bibinfo{title}{Entropy, Divergence, and Majorization in Classical and
  Quantum Thermodynamics}}, vol.~\bibinfo{volume}{16}
  (\bibinfo{publisher}{Springer Nature}, \bibinfo{year}{2022}).

\bibitem[{\citenamefont{Takahashi and Fujiwara}(2017)}]{Takahashi_001}
\bibinfo{author}{\bibfnamefont{K.}~\bibnamefont{Takahashi}} \bibnamefont{and}
  \bibinfo{author}{\bibfnamefont{A.}~\bibnamefont{Fujiwara}},
  \bibinfo{journal}{Journal of Physics A: Mathematical and Theoretical}
  \textbf{\bibinfo{volume}{50}}, \bibinfo{pages}{165301}
  (\bibinfo{year}{2017}),
  \urlprefix\url{https://dx.doi.org/10.1088/1751-8121/aa6326}.

\bibitem[{\citenamefont{Umegaki}(1962)}]{Umegaki_001}
\bibinfo{author}{\bibfnamefont{H.}~\bibnamefont{Umegaki}}, in
  \emph{\bibinfo{booktitle}{Kodai Mathematical Seminar Reports}}
  (\bibinfo{organization}{Department of Mathematics, Tokyo Institute of
  Technology}, \bibinfo{year}{1962}), vol.~\bibinfo{volume}{14}, pp.
  \bibinfo{pages}{59--85}.

\bibitem[{\citenamefont{Hiai and Petz}(2014)}]{Hiai_002}
\bibinfo{author}{\bibfnamefont{F.}~\bibnamefont{Hiai}} \bibnamefont{and}
  \bibinfo{author}{\bibfnamefont{D.}~\bibnamefont{Petz}},
  \emph{\bibinfo{title}{Introduction to matrix analysis and applications}}
  (\bibinfo{publisher}{Springer Science \& Business Media},
  \bibinfo{year}{2014}).

\bibitem[{\citenamefont{HIAI}(2010)}]{Hiai_007}
\bibinfo{author}{\bibfnamefont{F.}~\bibnamefont{HIAI}},
  \bibinfo{journal}{Interdisciplinary Information Sciences}
  \textbf{\bibinfo{volume}{16}}, \bibinfo{pages}{139} (\bibinfo{year}{2010}).

\bibitem[{\citenamefont{Bhatia}(1996)}]{Bhatia_001}
\bibinfo{author}{\bibfnamefont{R.}~\bibnamefont{Bhatia}},
  \emph{\bibinfo{title}{Matrix Analysis}}, Graduate Texts in Mathematics
  (\bibinfo{publisher}{Springer New York}, \bibinfo{year}{1996}), ISBN
  \bibinfo{isbn}{9780387948461},
  \urlprefix\url{https://books.google.co.jp/books?id=F4hRy1F1M6QC}.

\bibitem[{\citenamefont{Horn and Johnson}(2012)}]{Horn_001}
\bibinfo{author}{\bibfnamefont{R.}~\bibnamefont{Horn}} \bibnamefont{and}
  \bibinfo{author}{\bibfnamefont{C.}~\bibnamefont{Johnson}},
  \emph{\bibinfo{title}{Matrix Analysis}} (\bibinfo{publisher}{Cambridge
  University Press}, \bibinfo{year}{2012}), ISBN \bibinfo{isbn}{9780521548236},
  \urlprefix\url{https://books.google.co.jp/books?id=mlQ-PgAACAAJ}.

\end{thebibliography}

\end{document}